\documentclass[aps,prc,preprint,groupedaddress,showpacs]{revtex4}
\usepackage{epsfig}

\begin{document}

\title{Production of $\Theta^{+}$ in $\gamma +D \to \Lambda +\Theta^{+}$ and
$\gamma +D \to \Sigma +\Theta^{+}$ reactions}

\author{V. Guzey}

\affiliation{Institut f{\"u}r Theoretische Physik II, Ruhr-Universit{\"a}t
Bochum, D-44780 Bochum, Germany}

\begin{abstract}
The 
$\gamma +D \to \Lambda +\Theta^{+}$ and $\gamma +D \to \Sigma +\Theta^{+}$
reactions can be used to determine the width of $\Theta^+$ almost 
model-independently. 
We calculate the differential cross sections of the 
$\gamma +D \to \Lambda +\Theta^{+}$, $\gamma +D \to \Sigma +\Theta^{+}$ and
relevant background reactions in the photon energy range $1.2 \leq E_{\gamma} \leq 2.6$ GeV.
We determine the most favorable 
kinematic conditions and observables for the experimental studies of $\Theta^{+}$ in the 
considered processes. 
We argue that a comparison of the  $\gamma +D \to \Lambda +\Theta^{+}$ and 
$\gamma +D \to \Sigma +\Theta^{+}$ cross sections should unambiguously determine
isospin of $\Theta^+$.

\end{abstract}


\maketitle

\section{Introduction}
\label{sec:intro}

Evidence for the existence of the pentaquark state $\Theta^+$~\cite{Polyakov97} is now 
rather overwhelming.
The initial experimental reports about the discovery of 
$\Theta^+$~\cite{SPRING8,DIANA,CLAS,CLAS2,Neutrino,SAPHIR,HERMES,SVD,COSYTOF,ZEUS,GRAAL}
 will be followed by a series of dedicated high precision experiments aiming
to study such properties of the new baryon as parity, spin and isospin. 
Hence, it is topical to analyze in which reactions and at which 
kinematic conditions production of $\Theta^+$ is sufficiently copious  and
can be reliably estimated by hadronic phenomenology.

An important role in the present and future experimental investigations 
concerning $\Theta^+$ is played by photoproduction on deuterium.
In some cases, the deuterium target  serves as a mere source of the proton
and neutron 
targets and rescattering on the spectator nucleon simply enhances the  
$\Theta^+$ signal~\cite{CLAS}. In other cases, rescattering on the spectator
 nucleon
is the source of  $\Theta^+$ production~\cite{GRAAL}.
In this paper, we concern ourselves
with the latter class of reactions. In particular, 
we consider the strangeness tagging 
 $\gamma +D \to \Lambda (\Sigma) +\Theta^{+}$ reaction,
where the interaction of the photon with one of the nucleons produces a 
hyperon ($\Lambda$ or $\Sigma$) and a kaon. One then can choose kinematics
where the produced kaon has
the correct momentum to produce $\Theta^+$ via the interaction with the
 other nucleon of the deuteron.
The main advantage of the reaction is that the theoretical analysis 
is only weakly model-dependent: the differential cross section
for the $\gamma +p \to \Lambda (\Sigma)+K^+$ has been measured and
the $\gamma +p \to \Lambda(\Sigma) +K^+$ and $\gamma +n \to \Lambda (\Sigma) +K^0$
amplitudes have been 
phenomenologically parametrized;
  the deuteron wave function is known very well
for the momentum range involved in the process;
the dynamical information about $\Theta^{+}$ enters only through the
total width of $\Theta^{+}$, which can be treated as a free parameter
and then determined  by the comparison to experiment.
However, since production of  $\Theta^{+}$ takes place through rescattering
on the spectator nucleon of deuterium (see Fig.~\ref{fig:mech}), the
resulting  cross section is suppressed by the nuclear 
wave function at high momenta (this suppression is under control).

\section{$\gamma +D \to \Lambda (\Sigma) +\Theta^{+}$ cross sections}
\label{sec:main}

\begin{figure}[h]
\begin{center}
\epsfig{file=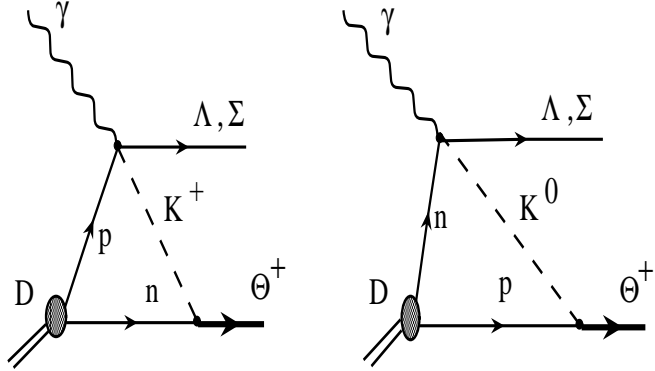,width=10cm,height=10cm}
\vskip -3cm
\caption{Two graphs of $\gamma +D \to \Lambda (\Sigma) +\Theta^{+}$.}
\label{fig:mech}
\end{center}
\end{figure}

The reaction $\gamma +D \to \Lambda (\Sigma) +\Theta^{+}$
is described by  two Feynman graphs presented in Fig.~\ref{fig:mech}.
We neglected the unknown final-state interactions between $\Lambda (\Sigma)$ and  $\Theta^{+}$.
It is important to note that the resulting scattering
 amplitude on deuterium  crucially depends on  interference between the 
$\gamma \, p \to \Lambda (\Sigma)  \, K^{+}$ and  
$\gamma \, n \to \Lambda (\Sigma) \, K^{0}$ (elementary) scattering amplitudes.
This introduces certain model-dependence into our calculations since
our results become dependent on the $\gamma \, n \to \Lambda (\Sigma) \, K^{0}$
amplitude, for which there is only a phenomenological parameterization and not
 experimental data.

Our numerical analysis of the Feynman graphs in Fig.~\ref{fig:mech} showed
that the scattering amplitude is predominantly imaginary in the considered
kinematics.
The imaginary part 
is found as a sum of all possible cuts of the diagrams in 
Fig.~\ref{fig:mech}. It is clear that there are three possible cuts. However, 
 cutting simultaneously the  spectator nucleon (neutron in the left graph and
proton in the right graph) and kaon lines
gives the dominant contribution. This cut places the spectator nucleon and the
kaon on mass shell.

After this cut,  in the $\Theta^+ \to N K$ vertex  
 all particles are on mass shell. This 
means that the corresponding expression is a 
function of the
 particle masses and the total width of $\Theta^+$ and, hence, it
does not depend on the spectator momentum.

The interacting nucleon (proton in left graph and
neutron in the right graph of Fig.~\ref{fig:mech}) is off mass shell. 
An examination of its energy denominator shows that the interacting nucleon
is not far from its mass shell and, hence, to a good approximation can be 
treated as being on mass shell. Then, with a good accuracy,
the $\gamma +N \to \Lambda + K$ amplitude in Fig.~\ref{fig:mech} 
depends only on the external four-momentum transfer 
$t=(p_{\gamma}-p_{\Lambda})^2$ and the photon energy $E_{\gamma}$.

Details of the derivation of the scattering amplitude and cross sections 
corresponding to Feynman graphs of Fig.~\ref{fig:mech} are given in 
Appendix~A. Here we give the final expression for the differential
 cross section for the $\gamma +D \to \Lambda (\Sigma) + \Theta^+$ process
\begin{equation}
\frac{d \sigma^{\gamma +D \to \Lambda (\Sigma)  + \Theta^+}}{dt}= 2 \pi \Gamma^{tot} \frac{M_{\Theta}^3}{\sqrt{(M_{\Theta}^2
-m^2-m_K^2)^2-4 m^2 m_K^2}} \frac{d \sigma^{p+n}}{dt} S(t) \,,
\label{eq:main1}
\end{equation}
where $\Gamma^{tot}$ is the total width of $\Theta^+$ and
$M_{\Theta}=1.540$ GeV is its mass; $m$ is the nucleon mass
and $m_K$ is the kaon mass.

The factorized form of Eq.~(\ref{eq:main1}) is a consequence of the 
on-mass-shellness of all particles in the loop in Fig.~\ref{fig:mech}.
The first factor involving the masses and $\Gamma^{tot}$
 comes from the $\Theta^+ \to N K$ vertex. 

We assumed that $\Theta^+$
has spin 1/2 and isospin 0. 
Other spin and isospin assignments are considered in the end of this Section.

\begin{figure}[h]
\begin{center}
\epsfig{file=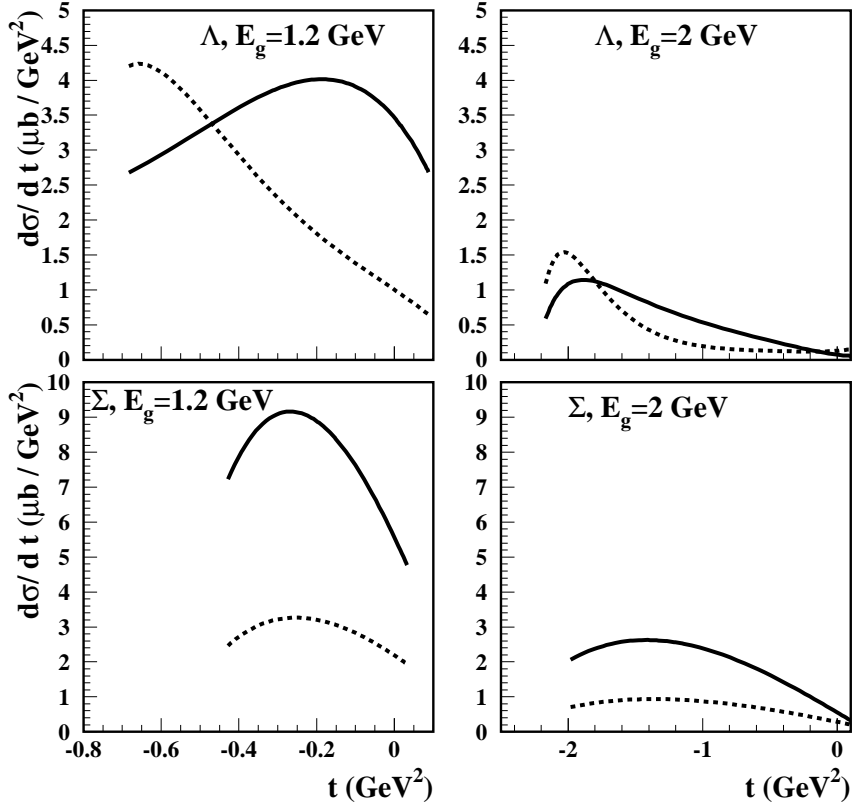,width=12cm,height=12cm}
\caption{The differential cross section $d \sigma^{p+n}/ dt$, which includes
both $\gamma +p \to \Lambda (\Sigma) + K^+$
and $\gamma +n \to \Lambda (\Sigma) + K^0$ amplitudes and their interference,
as a function of the momentum transfer squared $t$ from $\gamma$ to $\Lambda$  
 (solid curves). 
The $\gamma +p \to \Lambda (\Sigma) + K^+$ cross section $d \sigma^{p}/ dt$
is given by the dashed curves. The upper panels correspond to $\Lambda$ 
production, the lower panels correspond to $\Sigma$ 
production.}
\label{fig:risunok1}
\end{center}
\end{figure}

The second factor $d \sigma^{p+n}/ dt$ is the differential cross section,
which includes the 
$\gamma +p \to \Lambda (\Sigma) + K^+$
and $\gamma +n \to \Lambda (\Sigma) + K^0$ amplitudes and their interference.
The differential cross section $d \sigma^{p+n}/ dt$ 
at the photon beam energies 1.2 and 2 GeV
 is presented in Fig.~\ref{fig:risunok1} (solid curves). For comparison,
 we also give the $\gamma +p \to \Lambda (\Sigma) + K^+$ differential 
cross section (dashed curves).

Note that the momentum transfer squared from $\gamma$ to $\Lambda$ is 
defined as $t=(p_{\gamma}-p_{\Lambda})^2$. This means that in  
Fig.~\ref{fig:risunok1}, the minimal $t$ (maximal $|t|$) corresponds to the 
forward scattering kaon in the center of mass frame.

Note also the important role of interference between the 
$\gamma +p \to \Lambda (\Sigma) + K^+$
and $\gamma +n \to \Lambda (\Sigma) + K^0$ amplitudes.
At small values of $t$, which is the most interesting region for us,
$d \sigma^{p+n}/ dt$ is significantly
larger than $d \sigma^{p}/ dt$. Thus, the contribution
 of $\gamma +n \to \Lambda (\Sigma) + K^0$
 significantly enhances the resulting
$\gamma +D \to \Lambda(\Sigma)  + \Theta^+$ cross section.
The curves in Fig.~\ref{fig:risunok1} were obtained  using
the MAID data base and generator~\cite{MAID} consistent with the
SAPHIR $\gamma +p \to \Lambda + K^+$ data~\cite{SAPHIR:elem,SAPHIR:elemnew}
and  $\gamma +p \to \Sigma + K^+$ data~\cite{Sigma,SAPHIR:elemnew}.
Note that the choice of the photon energy $E_{\gamma}=1.2$ and 2 GeV 
is motivated 
by the constraints of the MAID generator, $0.9 < E_{\gamma} < 2.1$ GeV.  
Note also that
the cross section for production of $\Sigma$ is approximately two times larger than
that of $\Lambda$. 

It is important to note that since we took $\Theta^+$ with isospin-0 for this
calculation, the $\gamma +p \to \Lambda (\Sigma) + K^+$
and $\gamma +n \to \Lambda (\Sigma) + K^0$ amplitudes as generated by MAID 
enter with a relative plus sign.

\begin{figure}[h]
\begin{center}
\epsfig{file=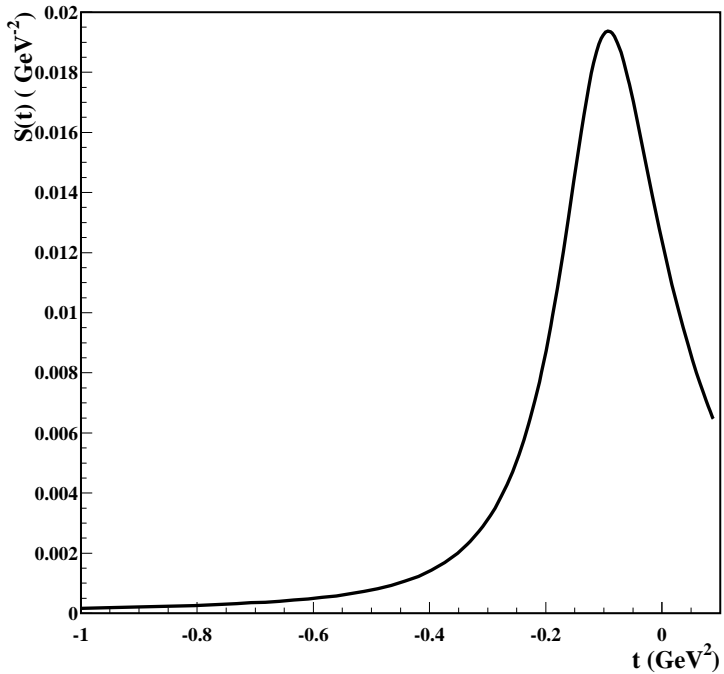,width=10cm,height=10cm}
\caption{The nuclear suppression factor $S(t)$.}
\label{fig:risunok2}
\end{center}
\end{figure}

The last factor in  Eq.~(\ref{eq:main1}) describes the $t$-dependent
suppression by the deuteron wave function due to rescattering
on the spectator nucleon (see Appendix~A for details).
We used the deuteron wave function with  the Paris nucleon-nucleon 
potential~\cite{Paris}.
Figure~\ref{fig:risunok2} shows
$S(t)$ as a function of $t$. Note that within our approach,
$S(t)$ does not depend on energy. Also, since masses of $\Lambda$ and $\Sigma$
are rather close, the nuclear factor $S(t)$ is the same for $\Lambda$ and
 $\Sigma$ production.
 
A very strong $t$-dependence of $S(t)$ suggests that 
the region 
$-0.2 < t < 0$ GeV$^2$ is most favorable for the experimental studies of 
the $\gamma +D \to \Lambda (\Sigma) + \Theta^+$ reaction.

\begin{figure}[t]
\begin{center}
\epsfig{file=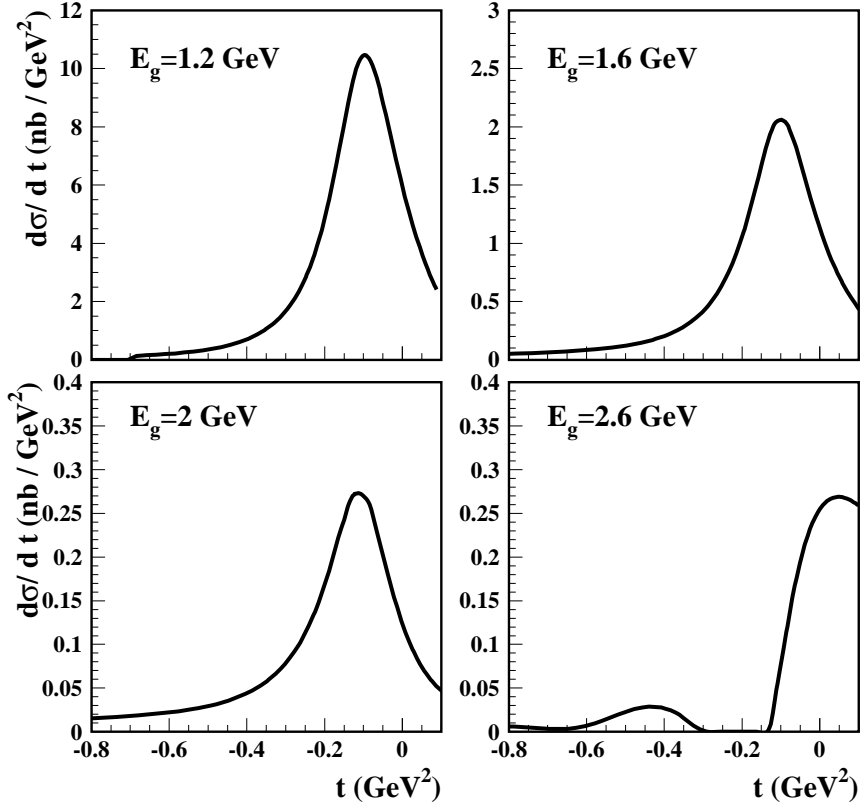,width=12cm,height=12cm}
\caption{The $\gamma +D \to \Lambda  + \Theta^+$ differential cross section
as a function of the momentum transfer $t$ at different photon energies, 
$E_{\gamma}=1.2$, 1.6, 2 and 2.6 GeV. The total width of $\Theta^+$ is 
assumed $\Gamma^{tot}=5$ MeV. In order to use a different width of $\Theta^+$, 
${\bar \Gamma}^{tot}$, the curves in this figure should be scaled by the factor
${\bar \Gamma}^{tot}/\Gamma^{tot}$.
}
\label{fig:risunok3}
\end{center}
\end{figure}

\begin{figure}[h]
\begin{center}
\epsfig{file=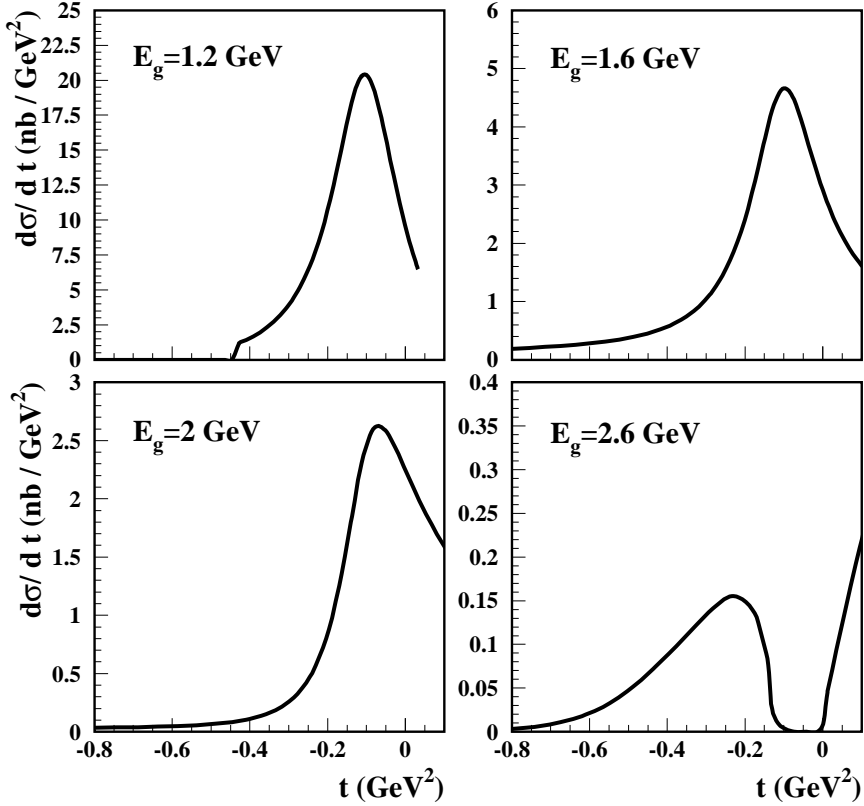,width=12cm,height=12cm}
\caption{The $\gamma +D \to \Sigma  + \Theta^+$ differential cross section
as a function of momentum transfer $t$ at different photon energies, 
$E_{\gamma}=1.2$, 1.6, 2 and 2.6 GeV. The total width of $\Theta^+$ is 
assumed $\Gamma^{tot}=5$ MeV.
In order to use a different width of $\Theta^+$, 
${\bar \Gamma}^{tot}$, the curves in this figure should be scaled by the factor
${\bar \Gamma}^{tot}/\Gamma^{tot}$.
}
\label{fig:risunok4}
\end{center}
\end{figure}

The main results of the present paper, the differential cross sections for
the $\gamma +D \to \Lambda + \Theta^+$ and $\gamma +D \to \Sigma + \Theta^+$
processes, are presented in Figs.~\ref{fig:risunok3} and \ref{fig:risunok4}.

The differential cross section is calculated at four different photon energies,
 $E_{\gamma}=1.2$, 1.6, 2 and 2.6 GeV, which correspond to the kinematics of
SPring-8, JLab, SAPHIR and GRAAL experiments (the photon energy $E_{\gamma}=2.6$ 
GeV is the largest energy
at which there are $\gamma +p \to \Lambda (\Sigma) + K^+$ data).
 One sees that the cross sections
are largest for $-0.2 < t < 0$ GeV$^2$: this is the most favorable region of 
$t$ for copious $\Theta^+$ production.
In this calculation we assumed that 
$\Gamma^{tot}=5$ MeV.
As follows from Eq.~(\ref{eq:main1}), the discussed cross sections depend linearly on 
$\Gamma^{tot}$. Therefore, if one wishes to use a different width of $\Theta^+$, 
${\bar \Gamma}^{tot}$, the cross section of $\Theta^+$ production
should be rescaled by the factor
${\bar \Gamma}^{tot}/\Gamma^{tot}$.

One should note that at the moment the  total width of $\Theta^+$
is rather uncertain. Theoretical predictions for its values range from
less than 15 MeV~\cite{Polyakov97} to several MeV~\cite{PWA,Cahn}.
The experimental determination of the width of $\Theta^+$ is limited
by the experimental resolution. The most stringent constraint is
given by the DIANA collaboration~\cite{DIANA}, $\Gamma^{tot} < 9$ MeV.
Only two experiments, HERMES~\cite{HERMES} and ZEUS~\cite{ZEUS}, 
appear to indicate a finite width of $\Theta^+$ which is somewhat
larger than the experimental resolution.

The $\gamma +D \to \Lambda (\Sigma) + \Theta^+$ 
cross section decreases rather rapidly with increasing photon energy.
This is determined by the decrease of the $\gamma +p \to \Lambda (\Sigma)+K^+$
cross section~\cite{SAPHIR:elemnew} and, more importantly, by 
the changing $t$-dependence of the
 $\gamma +p \to \Lambda (\Sigma)+K^+$ and $\gamma +n \to \Lambda (\Sigma)+K^0$
differential cross sections.
As the photon energy increases, the cross sections are peaked progressively
at smaller c.m. scattering angles of the kaon, which translates to
intermediate values of $t$ in our kinematics, where the suppression
 due to the nuclear factor 
$S(t)$ is rather significant.

~

One sees from Figs.~\ref{fig:risunok3} and \ref{fig:risunok4} that the 
photon energy $E_{\gamma}=1.2$ GeV gives the largest $\Theta^+$ production cross
section (notice the $y$-axis scale change as we increase the photon energy).

While for $E_{\gamma}=1.2$, 1.6 and 2  GeV we used the results of MAID~\cite{MAID},
for $E_{\gamma}=2.6$ GeV we directly used the recent SAPHIR experimental data for
$\gamma +p \to \Lambda (\Sigma)+K^+$~\cite{SAPHIR:elemnew}. Then we assumed
 that the $\gamma +p \to \Lambda (\Sigma)+K^+$ and $\gamma +n \to \Lambda (\Sigma)+K^0$
 scattering amplitudes are equal. Therefore, our results at $E_{\gamma}=2.6$ GeV
bear the largest theoretical uncertainty and should be considered as
a qualitative upper limit. This explains the shape of the curves at $E_{\gamma}=2.6$ GeV:
for instance, the presence of the dip at $t \approx -0.2$ GeV$^2$ in Fig.~\ref{fig:risunok3}
 originates from the almost zero $\gamma +p \to \Lambda +K^+$ cross section at 
$\cos(\Theta^{{\rm cms}}_{K^+}) \approx -0.25$~\cite{SAPHIR:elemnew}.

Integrating the differential cross sections over $t$, we obtain the 
corresponding total cross sections, which are summarized in 
Table~1.
\begin{table}[ht]
\begin{center}
\begin{tabular}{|c|c|c|}
\hline
$E_{\gamma}$, GeV & $\sigma^{\gamma +D \to \Lambda +\Theta^+}$, nbarn &
$\sigma^{\gamma +D \to \Sigma +\Theta^+}$, nbarn \\
\hline
1.2 & 2.51 & 4.44 \\
\hline
1.6 & 0.57 & 1.42  \\
\hline
2 & 0.090 & 0.74 \\
\hline
2.6 & 0.055 & 0.063 \\
\hline 
\end{tabular}
\caption{Integrated $\gamma +D \to \Lambda +\Theta^+$ and $\gamma +D \to \Sigma +\Theta^+$ cross sections at different photon energies $E_{\gamma}$
The total width of $\Theta^+$ is 
assumed $\Gamma^{tot}=5$ MeV.
}
\end{center}
\label{table:total}
\end{table}

One should note that in the considered processes, $\Lambda$ can be produced
 either directly in the $\gamma +N \to \Lambda +K$ vertex or in the
$\Sigma^0 \to \Lambda + \gamma$ decay. Hence, a special attention should be payed
to distinguish these two ways of producing $\Lambda$
in the experiment. 

Also, one can consider other than spin-1/2 and isospin-0 assignments for $\Theta^+$. The issue of isospin of $\Theta^+$
constitutes an interesting theoretical question, see 
for instance~\cite{Capstick,Page}. 
If $\Theta^+$ has spin-3/2, then the
final expression for the differential cross section in Eq.~(\ref{eq:a4})
 should be twice as large.
This is a consequence of the fact that spin-3/2 $\Theta^+$ has twice as many
 polarization states compared to the spin-1/2 $\Theta^+$.

\begin{figure}[h]
\begin{center}
\epsfig{file=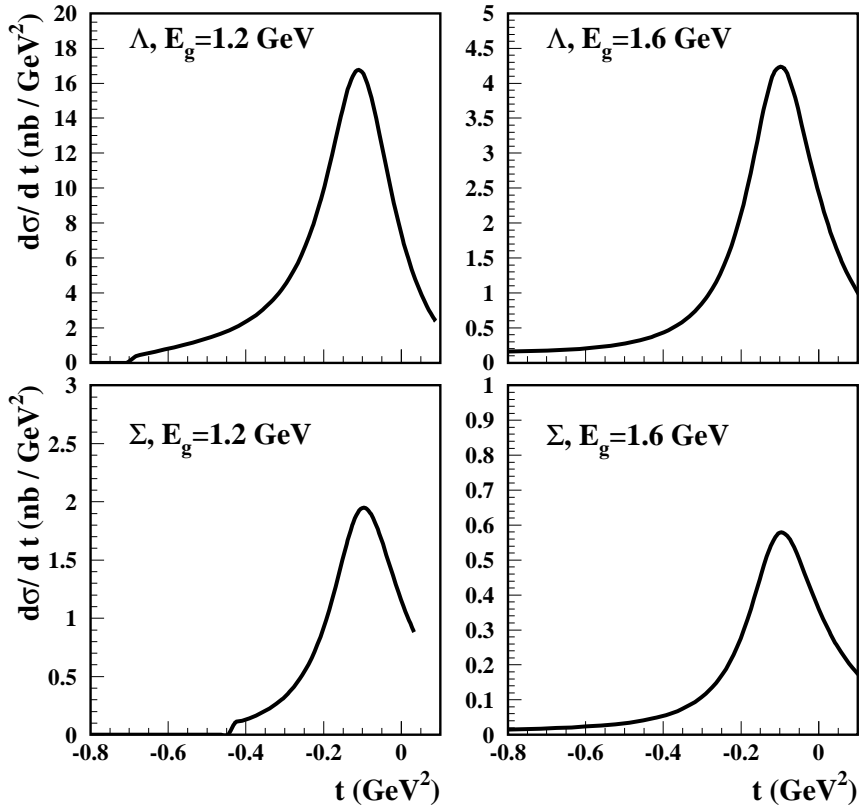,width=12cm,height=12cm}
\caption{$\Theta^+$ with isospin-1. The $\gamma +D \to \Sigma  + \Theta^+$ (upper panels) and 
$\gamma +D \to \Sigma^0 +\Theta^+$ (lower panels)
 differential cross sections
as functions of the momentum transfer $t$ at different photon energies, 
$E_{\gamma}=1.2$ and  1.6 GeV. It is assumed that $\Gamma^{tot}=5$ MeV.
}
\label{fig:risunok6}
\end{center}
\end{figure}

If $\Theta^+$ has isospin-1, this introduces a minus sign 
between the two Feynman graphs in Fig.~\ref{fig:mech}. Indeed, in this case 
isospin invariance indicates that the $\Theta^+ p K^0$ and $\Theta^+ n K^+$
vertices have opposite signs (see Appendix~B for details).
 As a result, the 
$\gamma +D \to \Lambda +\Theta^+$ cross section becomes enhanced, while
the  
$\gamma +D \to \Sigma^0 +\Theta^+$ cross section is significantly reduced,
as compared to the isospin-0 case.
The corresponding $\Theta^+$ isospin-1 differential cross sections at two photon energies, 
$E_{\gamma}=1.2$ and 1.6 GeV, are presented in Fig.~\ref{fig:risunok6}. 

Figure~\ref{fig:risunok6} ($I_{\Theta}=1$) should be compared to 
Figs.~\ref{fig:risunok4} and \ref{fig:risunok5} ($I_{\Theta}=0$).
In the former case, the $\gamma +D \to \Lambda +\Theta^+$ 
cross section is enhanced by approximately factor of two as compared to the 
$I_{\Theta}=0$ case. On the other hand, the 
$\gamma +D \to \Sigma^0 +\Theta^+$ 
cross section becomes reduced by approximately factor of ten as compared to
the $I_{\Theta}=0$ case. This result can be explained as follows.
If isospin of $\Theta^+$ is zero, the $\gamma +D \to \Lambda +\Theta^+$ reaction
involves the $I=0$ component of the initial photon, while 
$\gamma +D \to \Sigma +\Theta^+$ reaction involves the $I=1$ component of the photon.
An analogy with the vector meson dominance model suggests that the interaction mediated
by the $I=1$ component is stronger than by the $I=0$ component, exactly as we observe.
On the other hand, if isospin of $\Theta^+$ is 1, the situation is just the opposite and
production of $\Lambda$ is expected to be larger than production of $\Sigma$.


A comparison of the 
$\gamma +D \to \Lambda +\Theta^+$ and
$\gamma +D \to \Sigma^0 +\Theta^+$ rates can clearly distinguish
between the $\Theta^+$  with isospin-0 and isospin-1 scenarios.
Indeed, while the absolute value of the
cross sections is uncertain due to the uncertainty in
the total width of $\Theta^+$ as well as due to possible final state
interactions. However, the ratio of the
$\gamma +D \to \Lambda +\Theta^+$ and
$\gamma +D \to \Sigma^0 +\Theta^+$
cross sections (let us denote it by $R$) 
is insensitive to $\Gamma^{tot}$ which cancels in the ratio. In addition,
the final state interaction is a correction and, hence, cannot change
$R$ too much. Since the ratio $R$ changes from $R \approx 0.5$ at $t \approx -0.1$ GeV$^2$
 (if $\Theta^+$ has isospin-0)
to $R \approx 8$ (in the isospin-1 case), it should be possible to experimentally 
distinguish  between these two cases.

If one assumes that $I_{\Theta}=1$, in addition to the $\Sigma^0 \Theta^+$ final state, 
one can consider production
 of $\Sigma^+ \Theta^0$ and $\Sigma^- \Theta^{++}$ states.
It follows from isospin conservation that 
$\sigma(\gamma +D \to \Sigma^+ +\Theta^0)=\sigma(\gamma +D \to \Sigma^- +\Theta^{++})$.
 However, based solely on isospin conservation, 
$\sigma(\gamma +D \to \Sigma^+ +\Theta^0)$ and $\sigma(\gamma +D \to \Sigma^- +\Theta^{++})$
 cannot be related to 
$\sigma(\gamma +D \to \Sigma^0 +\Theta^+)$.

We also note that the $\Theta^+$ production mechanism of Fig.~\ref{fig:mech} cannot produce 
$\Theta^+$ with isospin-2.

\section{Background estimates, interference with signal}

\begin{figure}[h]
\begin{center}
\epsfig{file=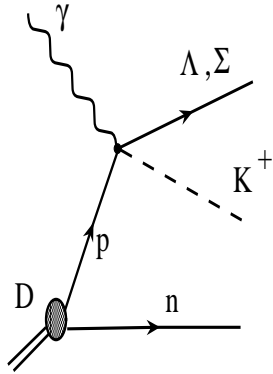,width=10cm,height=10cm}
\vskip -3cm
\caption{The dominant background reaction.}
\label{fig:bg}
\end{center}
\end{figure}

The main background reaction $\gamma +D \to \Lambda (\Sigma) + K^+ +n$ 
is presented in Fig.~\ref{fig:bg}. Since the other background reaction
 $\gamma +D \to \Lambda (\Sigma) + K^0 +p$ involves the unmeasured
$\gamma +n \to \Lambda (\Sigma) + K^0$ amplitude, we shall concentrate
on the process in Fig.~\ref{fig:bg}.

The background process plays a two-fold role. First, a detailed examination shows
that interference with the signal enhances $\Theta^+$ production (see Appendix~B for details).
 The differential interference cross section can be cast in the form of
 Eq.~(\ref{eq:main1})
\begin{equation}
\frac{d \sigma^{I}}{dt}= 2 \pi \Gamma^{tot} \frac{M_{\Theta}^3}{\sqrt{(M_{\Theta}^2
-m^2-m_K^2)^2-4 m^2 m_K^2}} \frac{d \sigma^{I}_{{\rm elem}}}{dt} S(t) \,,
\label{eq:bg1}
\end{equation}
where $\sigma^{I}_{{\rm elem}}$ is the interference cross section obtained using the interference amplitude
$|{\cal A}^I_{{\rm elem}}|^2=({\cal A}^{p}+{\cal A}^{n})\left({\cal A}^{p} \right)^{\ast}+
\left({\cal A}^{p}+{\cal A}^{n}\right)^{\ast} {\cal A}^{p}$ involving the
$\gamma +p \to \Lambda (\Sigma) + K^+$ (denoted as ${\cal A}^{p}$) and 
$\gamma +n \to \Lambda (\Sigma) + K^0$ (denoted as ${\cal A}^{n}$) amplitudes.

Hence, the signal plus interference cross section takes the following compact form
\begin{equation}
\frac{d \sigma^{\gamma+D \to \Lambda (\Sigma)+n+K^+}}{dt}= 2\pi \Gamma^{tot} \frac{M_{\Theta}^3}{\sqrt{(M_{\Theta}^2
-m^2-m_K^2)^2-4 m^2 m_K^2}} \left(\frac{1}{2}\frac{d \sigma^{p+n}}{dt}+\frac{d \sigma^{I}_{{\rm elem}}}{dt} \right) S(t) \,,
\label{eq:bg2}
\end{equation}
where an additional factor 1/2 in front of the first term is a consequence of the 1/2-branching ratio
of the $\Theta^+ \to n K^+$ decay. Note that the simple form of Eq.~(\ref{eq:bg2}) holds only for
the cross sections integrated over the final neutron momentum (see discussion below).

The second aspect of the background process is that it produces the same final state as the signal
process but does not carry any information (unlike the interference term) about $\Theta^+$.
 Hence, we 
intend to find kinematic conditions where the signal plus interference cross section is larger
than the purely background contribution.
It is important to point out that in general, the purely background (Born) cross section of
Fig.~\ref{fig:bg} is much larger than the rescattering cross section of 
Fig.~\ref{fig:mech}. The relative magnitude of these cross sections is  determined
predominantly by the nuclear suppression factor.
While the Born amplitude is
proportional to $\psi_D(p_s)$ with $p_s$ being the final nucleon (spectator) momentum,
the rescattering amplitude involves $\int dk k \psi_D(k)$.
Therefore, one way to suppressed the purely background cross section is to choose 
relatively large spectator (neutron in Fig.~\ref{fig:bg}) momenta.

Thus, it is important to discuss how different contributions depend on $p_s$. The 
$p_s$-dependence of the signal cross section is rather weak and is determined by the
neutron-kaon phase space. The interference term depends on $p_s$ much stronger through
$\psi_D(p_s)$. In addition, the interference term is suppressed by the smallness of the
total width of $\Theta^+$ and the loop integral involving the deuteron wave function.
The $p_s$-dependence of the
purely background contribution is the strongest: the cross section is proportional to
$|\psi_D(p_s)|^2$. However, this suppression is still not sufficient for the reduction
of the background to the signal level: One needs to impose an additional constraint
on the neutron-kaon invariant mass, see Eq.~(\ref{eq:bg}).

The theoretical analysis of the Feynman diagram in Fig.~\ref{fig:bg} is rather
straightforward and is detailed in Appendix~B. Since we aim to estimate the
level of background under the $\Theta^+$ resonance peak in the
$K^+$-neutron system, we impose the condition that in Fig.~\ref{fig:bg}, 
the invariant mass of the $K^+$-neutron system, $M_{K^+\,n}$, belongs to 
the interval, for instance, 
\begin{equation}
M_{\Theta}-10 \, {\rm MeV} \leq M_{K^+\,n} \leq M_{\Theta}+10 \, {\rm MeV} \,.
\label{eq:bg}
\end{equation}
The value of 10 MeV is chosen as a realistic example with only one constraint in mind that
 10 MeV resolution is clearly within the reach of future 
dedicated experiments on $\Theta^+$ production.

\begin{figure}[h]
\begin{center}
\epsfig{file=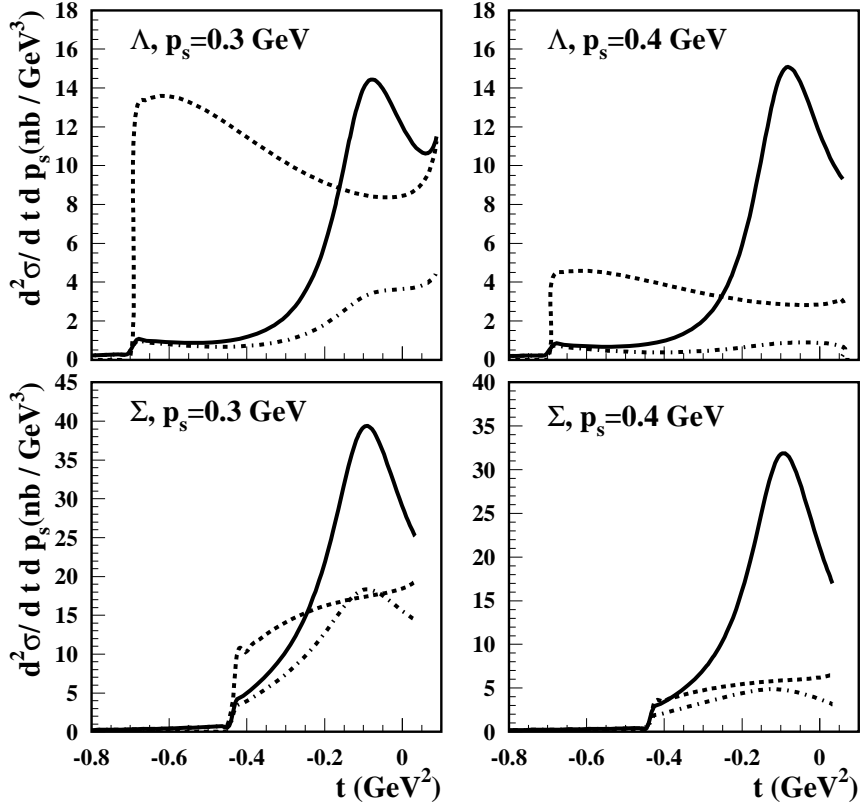,width=12cm,height=12cm}
\caption{The $\gamma +D \to \Lambda (\Sigma)  + n+K^+$ double differential
 cross sections as functions of $t$ at different spectator momenta $p_s$.
The solid curves represent the signal plus interference cross section;
the dash-dotted curves give the interference cross section;
the dashed curves give the purely background cross section.
The total width of $\Theta^+$ is 
assumed $\Gamma^{tot}=5$ MeV; the photon energy $E_{\gamma}=1.2$ GeV.}
\label{fig:risunok5}
\end{center}
\end{figure}

The purely background $\gamma +D \to \Lambda (\Sigma) + K^+ +n$
double differential cross sections corresponding to   Fig.~\ref{fig:bg} 
at different values of the momentum of the spectator neutron
are presented in  Fig.~\ref{fig:risunok5}
by dashed curves. They should be compared to solid curves, which represent
the signal plus interference
$\gamma +D \to \Lambda (\Sigma) + K^+ +n$ cross section (see Appendix~B for
the exact expressions).
The interference cross section is given by the dot-dashed curves. Note that
interference with the background increases the signal.
All curves correspond to $E_{\gamma}=1.2$ GeV,
where $\Theta^+$ production is largest.

As one can see from Fig.~\ref{fig:risunok5}, choosing a sufficiently high 
spectator momentum, $p_s=300$ and $400$ MeV/c and above, significantly reduces
the background without changing the position and shape of the resonance peak.
Hence, detection of a spectator nucleon (neutron) in coincidence with
$\Lambda (\Sigma)$ presents a good opportunity to increase the signal to 
background ratio under the $\Theta^+$ peak in the $\Theta^+$ photoproduction 
on deuterium.

\begin{figure}[h]
\begin{center}
\epsfig{file=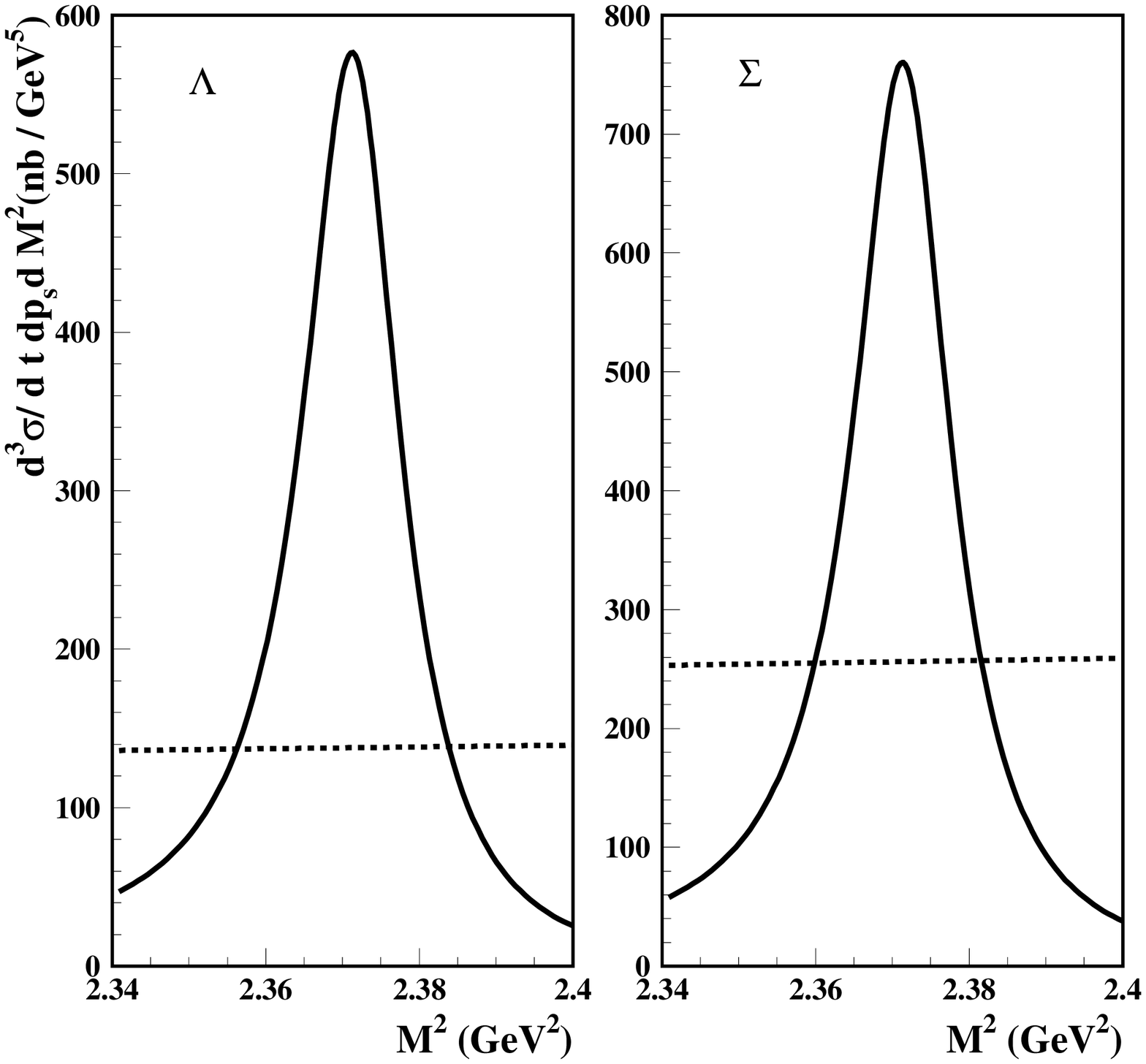,width=10cm,height=10cm}
\caption{The $\gamma +D \to \Lambda (\Sigma)  + n+K^+$ triple differential
 cross sections as functions of the proton-kaon invariant mass squared at
$E_{\gamma}=1.2$ GeV, $t=-0.1$ GeV$^2$ and $p_s=300$ MeV/c.
The solid curves represent the signal plus interference cross section,
the dashed curves give the purely background cross section.
}
\label{fig:risunok7}
\end{center}
\end{figure}

If experimental resolution in the final neutron-kaon invariant mass is at the level of
several MeV and luminosity is sufficiently high, one can also study the shape of the 
$\Theta^+$ 
production cross section
as a function of $M^2=(p_n+p_K)^2$ and clearly separate the signal from the background.
An example is presented in Fig.~\ref{fig:risunok7}, where we plot the signal plus interference
triple differential cross section (solid curves) and the purely background triple
 differential cross section
(dashed curves) as functions of $M^2$. The kinematics is maximally favorable for the
 signal extraction:
$E_{\gamma}=1.2$ GeV, $t=-0.1$ GeV$^2$ and $p_s=300$ MeV/c.
One can see in Fig.~\ref{fig:risunok7} that the distinctive Breit-Wigner curve of the
 signal plus 
interference (solid curves) dominates the flat background distribution (dashed
curves).

\section{Conclusions and Discussion}

Among a multitude of $\Theta^+$ photoproduction mechanisms, we single out the
$\gamma +D \to \Lambda (\Sigma)+ \Theta^+$ process as being weakly model-dependent.
The $\gamma +D \to \Lambda (\Sigma)+ \Theta^+$ cross section
involves the measured $\gamma +p \to \Lambda (\Sigma)+ K^+$ and phenomenologically
parametrized $\gamma +n \to \Lambda (\Sigma)+ K^0$ amplitudes,
the well established deuteron wave function and the total width of $\Theta^+$.
Therefore, the $\gamma +D \to \Lambda (\Sigma)+ \Theta^+$ process is a very good candidate
reaction for a precise measurement of the $\Theta^+$ width.

We studied the $\gamma +D \to \Lambda (\Sigma)+ \Theta^+$ differential cross section and
found it to be sharply peaked for $-0.2 < t < 0$ GeV$^2$, where $t=(p_{\gamma}-p_{\Lambda})^2$.
Hence, this kinematic region is most favorable for the experimental studies.

We performed our analysis of the $\gamma +D \to \Lambda (\Sigma)+ \Theta^+$ cross section
for a range of the photon energies, $1.2 \leq E_{\gamma} \leq 2.6$ GeV, which almost covers
the energy range of SPring-8, TJNAF, SAPHIR and GRAAL experiments. It is found that
$\gamma +D \to \Lambda (\Sigma)+ \Theta^+$ cross section is largest at the smallest studied
energy, $E_{\gamma}=1.2$ GeV.

In order to understand if the extraction of $\Theta^+$ from the 
$\gamma +D \to \Lambda (\Sigma)+ \Theta^+$ data 
is possible, we estimated the rate of the background reaction
$\gamma +D \to \Lambda (\Sigma)+ K^+ + n$. We showed that the signal to background ratio
is large when one chooses sufficiently large momenta of the spectator neutron, 
$p_s > 300$ MeV/c.

In addition, if experimental accuracy in the determination of the final neutron-kaon invariant mass
is at the level of several MeV, one can also study the $\Theta^+$ photoproduction cross section
as a function of that invariant mass. The shape of the resulting distribution will clearly
separate the signal from the background, see Fig.~\ref{fig:risunok7}. 

Our calculations in this paper are based on the assumption that $\Theta^+$ has spin-1/2 and 
isospin-0.
If $\Theta^+$ is assigned spin-3/2, then the signal and interference cross sections are twice as large.
The relative rates of $\Lambda \Theta^+$ and $\Sigma \Theta^+$ production are very sensitive to isospin of 
$ \Theta^+$. If $\Theta^+$ has isospin-0, production of $\Sigma$ is approximately twice as
large as production of $\Lambda$. However,
if we suppose that $\Theta^+$ has isospin-1, production of 
$\Sigma$ is significantly smaller than production of $\Lambda$.

Another important and yet undetermined characteristics of $\Theta^+$ is its parity.
 Our prediction for the $\gamma +D \to \Lambda (\Sigma)+ \Theta^+$ differential cross section does
 not depend on the $\Theta^+$ parity since $\Theta^+$
enters into our analysis only through its mass and total width. 
Overall, it appears that the $\gamma +D \to \Lambda (\Sigma)+ \Theta^+$ reaction does not seem to be a 
good candidate to study parity of $\Theta^+$. In general, the  $\gamma +D \to \Lambda (\Sigma)+ \Theta^+$ amplitude
can be parametrized in terms of twelve functions. Since the number of unknown functions is rather large,
choosing different polarizations of the photon, deuteron and $\Lambda (\Sigma)$ is not sufficient
 to produce unambiguous and model-independent relations between polarization observables and parity of
$\Theta^+$, as was suggested for the $\gamma +N \to {\bar K}+\Theta^+$ reaction~\cite{Rekalo}.

\section*{Acknowledgments}

It is a pleasure to thank Mark Strikman and Maxim Polyakov for continuing interest and illuminating discussions concerning the  $\gamma +D \to \Lambda (\Sigma)+ \Theta^+$ reactions. We also thank Kazuo Tsushima for 
the discussion of initial drafts of the 
present paper, as well as K. Goeke and V. Metag.
This
work is supported by the Sofia Kovalevskaya Program of the Alexander
von Humboldt Foundation.

\section*{Appendix A}

In this appendix, we derive the master equation for the $\gamma +D \to \Lambda (\Sigma) + \Theta^+$ cross section, Eq.~(\ref{eq:main1}).
The scattering amplitude corresponding to either one of the Feynman graphs of 
Fig.~\ref{fig:mech} reads
\begin{equation}
{\cal A}=i \int \frac{d^4 k}{(2 \pi)^4} \bar{u}(p_{\Theta}) \hat{\Gamma}_{\Theta}
\frac{\hat{k}+m}{k^2-m^2+i0} \frac{1}{(p_{\Theta}-k)^2-m_K^2+i0} 
\bar{u}(p_{\Lambda}) \hat{\Gamma}_{\Lambda} 
\frac{\hat{p}_D-\hat{k}+m}{(p_D-k)^2-m^2+i0} \hat{\Gamma}_{D} \,,
\label{eq:a1}
\end{equation}
where $k$ is the momentum of the spectator nucleon; the $\hat{\Gamma}_{\Theta}$
vertex describes the $\Theta^+ \to NK$ transition; the $\hat{\Gamma}_{\Lambda}$
vertex describes the $\gamma + N \to \Lambda (\Sigma) +K$ transition;
 $\hat{\Gamma}_{D}$ describes the $D \to NN$ transition.
For brevity, all spin polarization indices are implicit.

An explicit calculations shows that 
the imaginary part dominates the scattering amplitude in Eq.~(\ref{eq:a1}). 
The imaginary part is found by taking all possible cuts of the graphs in 
Fig.~\ref{fig:mech}. Cutting through the spectator nucleon and the kaon gives
the principal contribution and the resulting amplitude reads
\begin{eqnarray}
Im {\cal A}&=&\frac{1}{16 \pi}\int \frac{d k k}{E} \frac{\Theta(E_{\Theta}-E)}{p_{\Theta}} \Theta\left(-1 < \frac{E_{\Theta}E-a}{p_{\Theta}k} < 1\right)
 \bar{u}(p_{\Theta}) \hat{\Gamma}_{\Theta} u(k) \nonumber\\
&&\times
\bar{u}(p_{\Lambda}) \hat{\Gamma}_{\Lambda} u(p_D-k)
\frac{1}{(p_D-k)^2-m^2+i0} \bar{u}(k) \bar{u}(p_D-k)
 \hat{\Gamma}_{D} \,,
\label{eq:a2}
\end{eqnarray}
where $E_{\Theta}$ and $p_{\Theta}$ is the energy and momentum of $\Theta^+$
in the deuteron rest frame;
$a=(M_{\Theta}^2+m^2-m_K^2)/2$. 

In the $\Theta^+ \to N K$ vertex  
 all particles are on mass shell. Therefore, $\bar{u}(p_{\Theta}) \hat{\Gamma}_{\Theta} u(k)$ does not depend on the momentum $k$ and can be expressed solely
through the masses of $\Theta^+$, kaon and nucleon and
the total width of $\Theta^+$.

The $D \to NN$ vertex $\hat{\Gamma}_{D}$ is expressed through the non-relativistic deuteron wave function $\psi_D$
\begin{equation}
\frac{1}{(p_D-k)^2-m^2+i0} \bar{u}(k) \bar{u}(p_D-k)
 \hat{\Gamma}_{D}=\sqrt{(2 \pi)^3 2m}\, \psi_D(k) \,
\label{eq:a3}
\end{equation}
where the deuteron polarization is implicit.

Finally, since the nuclear wave function has a very strong dependence on the 
momentum $k$, the elementary amplitude 
$\bar{u}(p_{\Lambda}) \hat{\Gamma}_{\Lambda} u(p_D-k)$
can be taken out of integration at some average momentum $\langle k \rangle$.
It is important to emphasize that the resulting amplitude is a sum of both
graphs in Fig.~\ref{fig:mech} such that it involves the coherent sum of the
amplitudes $\gamma + p \to \Lambda (\Sigma^0) K^+$ and 
$\gamma + n \to \Lambda (\Sigma^0) K^0$.

The resulting spin-averaged differential cross section takes the following 
factorized form
\begin{equation}
\frac{d \sigma^{\gamma +D \to \Lambda (\Sigma)  + \Theta^+}}{dt}= 2 \pi \Gamma^{tot} \frac{M_{\Theta}^3}{\sqrt{(M_{\Theta}^2
-m^2-m_K^2)^2-4 m^2 m_K^2}} \frac{d \sigma^{p+n}}{dt} S(t) \,,
\label{eq:a4}
\end{equation}
where $\Gamma^{tot}$ is the total width of $\Theta^+$. 

This derivation assumed that $\Theta^+$ has spin-1/2. If we assumed that it has spin-3/2,
the final expression for the differential cross section in Eq.~(\ref{eq:a4}) would be twice
as large. This is a consequence of the fact that spin-3/2 $\Theta^+$ has twice as many
 polarization states compared to the spin-1/2 $\Theta^+$.

The differential cross section $d \sigma^{p+n} /dt$ involves the sum of the
$\gamma + p \to \Lambda (\Sigma^0) K^+$ amplitude (denoted ${\cal A}^p$)
 and the
$\gamma + n \to \Lambda (\Sigma^0) K^0$ amplitude (denoted ${\cal A}^n$)
\begin{equation}
\frac{d \sigma^{p+n}}{dt}=\frac{1}{64 \pi (E_{\gamma} m)^2} |{\cal A}^p+{\cal A}^n|^2 \,.
\label{eq:a5}
\end{equation}
Both amplitudes are taken as generated by the MAID generator~\cite{MAID} except for the case
$E_{\gamma}=2.6$ GeV, when we assumed that ${\cal A}^p={\cal A}^n$ and took
$|{\cal A}^p|^2$ from the experimental data~\cite{SAPHIR:elemnew}.

Note that if $\Theta^+$ has isospin-0, the $\gamma + p \to \Lambda (\Sigma^0) K^+$
and $\gamma + n \to \Lambda (\Sigma^0) K^0$ amplitudes should be added.
 However, if $\Theta^+$ has isospin-1,
the  $\gamma + p \to \Lambda (\Sigma^0) K^+$
and $\gamma + n \to \Lambda (\Sigma^0) K^0$ amplitudes should be subtracted, which leads
to an enhancement of the  $\gamma+D \to \Lambda +\Theta^+$ cross section and
a significant suppression of the $\gamma+D \to \Sigma^0 +\Theta^+$ cross section 
(see Fig.~\ref{fig:risunok6}).

The factor $S(t)$ describes the suppression due to the nuclear wave function
\begin{eqnarray}
S(t)&=&\left(\frac{\sqrt{(2 \pi)^3 2m}}{16 \pi} \right)^2 \int \frac{d k_1 k_1}{E_1} \frac{d k_2 k_2}{E_2}  
\frac{\Theta(E_{\Theta}-E_1)}{p_{\Theta}}\frac{\Theta(E_{\Theta}-E_2)}{p_{\Theta}}  \nonumber\\
&& \times \Theta\left(-1 < \frac{E_{\Theta}E_1-a}{p_{\Theta}k_1} < 1\right)
\Theta\left(-1 < \frac{E_{\Theta}E_2-a}{p_{\Theta}k_2} < 1\right)
 \rho_D(k_1,k_2) \,,
\label{eq:a6}
\end{eqnarray}
where $\rho_D(k_1,k_2)$ is the unpolarized deuteron density matrix, 
which can be expressed in terms of the $S$ (its wave function is denoted by $u(k)$) and $D$ (its wave function
is denoted by $w(k)$) waves
 of the deuteron wave function
\begin{equation}
\rho_D(k_1,k_2)=u(k_1)u(k_2) +w(k_1)w(k_2) \left(\frac{3}{2} \frac{({\vec k}_1 \cdot {\vec k}_2)^2}{k_1^2 
k_2^2}-\frac{1}{2} \right) \,.
\end{equation}
We used the Paris nucleon-nucleon potential for the deuteron wave function~\cite{Paris}.

\section*{Appendix B}

In this appendix we give expressions for the
interference and purely background cross sections.
We also present explicit expressions for the double and triple 
differential cross sections
plotted in Figs.~\ref{fig:risunok5} and \ref{fig:risunok7}.

The scattering amplitude for the background reaction in Fig.~\ref{fig:bg}
 reads
\begin{equation}
{\cal A}^{{\rm BG}}=-\bar{u}(p_{\Lambda}) \hat{\Gamma}_{\Lambda}^p 
\frac{\hat{p}_D-\hat{k}+m}{(p_D-k)^2-m^2+i0} \bar{u}(p_s) \hat{\Gamma}_{D} \,,
\label{eq:b1}
\end{equation}
which, after the non-relativistic reduction of $\hat{\Gamma}_{D}$
(see Eq.~(\ref{eq:a3})), becomes
\begin{equation}
{\cal A}^{{\rm BG}}=-\sqrt{(2 \pi)^3 2m}\, \bar{u}(p_{\Lambda}) \hat{\Gamma}_{\Lambda}^p 
u(p_D-p_s) \psi_{D}(p_s) \,.
\label{eq:b2}
\end{equation}
This amplitude interferes with the signal amplitude (see also 
Eq.~(\ref{eq:a2}))
\begin{eqnarray}
{\cal A}^{{\rm Signal}}&=&-i \int \frac{d^4 k}{(2 \pi)^4}
\bar{u}(p_s) {\hat \Gamma}_{\Theta} \frac{\hat{p}_{\Theta}+M_{\Theta}}{p_{\Theta}^2-M_{\Theta}^2+i \Gamma^{tot}M_{\Theta}} \hat{\Gamma}_{\Theta}
\frac{\hat{k}+m}{k^2-m^2+i0} \frac{1}{(p_{\Theta}-k)^2-m_K^2+i0} \nonumber\\
&& \times \bar{u}(p_{\Lambda}) \hat{\Gamma}_{\Lambda}^{p+n} 
\frac{\hat{p}_D-\hat{k}+m}{(p_D-k)^2-m^2+i0} \hat{\Gamma}_{D} \,.
\label{eq:b3}
\end{eqnarray}
Keeping only the imaginary part of the loop integral (which is a dominant contribution)
  and performing the non-relativistic
reduction (see Eq.~(\ref{eq:a3})), the amplitude in Eq.~(\ref{eq:b2}) becomes
\begin{eqnarray}
{\cal A}^{{\rm Signal}}&=&-i\left(\frac{\sqrt{(2 \pi)^3 2m}}{16 \pi} \right) 
\frac{1}{16 \pi}\int \frac{d k k}{E} \frac{\Theta(E_{\Theta}-E)}{p_{\Theta}} \Theta\left(-1 < \frac{E_{\Theta}E-a}{p_{\Theta}k} < 1\right)\nonumber\\
&& \times \bar{u}(p_s) {\hat \Gamma}_{\Theta} \frac{\hat{p}_{\Theta}+M_{\Theta}}{p_{\Theta}^2-M_{\Theta}^2+i \Gamma^{tot}M_{\Theta}} \hat{\Gamma}_{\Theta}  u(k) 
\bar{u}(p_{\Lambda}) \hat{\Gamma}_{\Lambda}^{p+n} u(p_D-k) \psi_D(k)  \,.
\label{eq:b4}
\end{eqnarray}
Near the resonance $p_{\Theta}^2-M_{\Theta}^2 \ll \Gamma^{tot}M_{\Theta}$ and, hence, 
 the signal 
amplitude is predominantly real. Therefore,  ${\cal A}^{{\rm Signal}}$ interferes
with ${\cal A}^{{\rm BG}}$ and this interference is constructive (both amplitudes 
have a negative sign in front of them). 

The interference cross section is obtained using Eqs.~(\ref{eq:b2}) and 
(\ref{eq:b4}) and can be cast in the form of Eq.~(\ref{eq:main1})
\begin{equation}
\frac{d \sigma^{I}}{dt}= 2 \pi \Gamma^{tot} \frac{M_{\Theta}^3}{\sqrt{(M_{\Theta}^2
-m^2-m_K^2)^2-4 m^2 m_K^2}} \frac{d \sigma^{I}_{{\rm elem}}}{dt} S(t) \,,
\label{eq:b5}
\end{equation}
where 
\begin{equation}
\frac{d \sigma^{I}_{{\rm elem}}}{dt}=\frac{1}{64 \pi (E_{\gamma} m)^2}\left(({\cal A}^{p}+{\cal A}^{n})\left({\cal A}^{p} \right)^{\ast}+
\left({\cal A}^{p}+{\cal A}^{n}\right)^{\ast} {\cal A}^{p} \right) \,.
\label{eq:b6}
\end{equation}

While the signal and interference cross sections have a similar appearance, the double differential signal
and interference cross sections have distinctly different dependences on the momentum of the neutron 
in the final state $p_s$. When we 
take into account the decay of $\Theta^+$ into the $n K^+$ final state,
Eq.~(\ref{eq:main1}) should be multiplied
 by the branching ratio of the  $\Theta^+ \to n K^+$ decay (which equals 1/2) and by the 
$n K^+$ phase space. The $\gamma +D \to \Lambda (\Sigma) + \Theta^+ \to
\Lambda (\Sigma) + n + K^+$ double differential cross section then reads
\begin{eqnarray}
\frac{ d^2 \sigma}{dt\, dp_s}&=&2 \pi \Gamma^{tot} \frac{M_{\Theta}^3}{\sqrt{(M_{\Theta}^2
-m^2-m_K^2)^2-4 m^2 m_K^2}} \frac{d \sigma^{p+n}}{dt} S(t) \nonumber\\
 && \times \frac{1}{2} \frac{M_{\Theta}}{2 p_{\Theta} k^{\ast}} \frac{p_s}{E_s} \Theta\left(-1 < \frac{m_K^2+p_{\Theta}^2+p_s^2-(E_{\Theta}-E_s)^2}{2 p_{\Theta} p_s} <1 \right) \,,
\label{eq:b7}
\end{eqnarray}
where $k^{\ast}$ is the spectator momentum in the $\Theta^+$ rest frame.

The $p_s$ dependence of the interference cross section is determined by the deuteron wave function and
the double differential cross section reads
\begin{equation}
\frac{d^2 \sigma^{I}}{dt\, dp_s}= 2 \pi \Gamma^{tot} \frac{M_{\Theta}^3}{\sqrt{(M_{\Theta}^2
-m^2-m_K^2)^2-4 m^2 m_K^2}} \frac{d \sigma^{I}_{{\rm elem}}}{dt} \frac{d S(t,p_s)}{d p_s} \,,
\label{eq:b8}
\end{equation}
where (see also Eq.~(\ref{eq:a6}))
\begin{eqnarray}
\frac{d S(t,p_s)}{d p_s}&=&\left(\frac{\sqrt{(2 \pi)^3 2m}}{16 \pi} \right)^2 \int \frac{d k k}{E_k} \frac{p_s}{E_s}  
\frac{\Theta(E_{\Theta}-E_k)}{p_{\Theta}}\frac{\Theta(E_{\Theta}-E_s)}{p_{\Theta}}  \nonumber\\
&& \times \Theta\left(-1 < \frac{E_{\Theta}E_k-a}{p_{\Theta}k} < 1\right)
\Theta\left(-1 < \frac{E_{\Theta}E_s-a}{p_{\Theta}p_s} < 1\right)
 \rho_D(k,p_s) \,.
\label{eq:b9}
\end{eqnarray}
The sum of Eqs.~(\ref{eq:b7}) and (\ref{eq:b9}) is presented in Fig.~\ref{fig:risunok5} by solid curves.
The interference cross section of Eqs.~(\ref{eq:b9}) is presented by dash-dotted curves in Fig.~\ref{fig:risunok5}.

The double differential cross section of the purely background process based on the amplitude of
 Eq.~(\ref{eq:b2}) takes the form
\begin{equation}
\frac{ d^2 \sigma^{{\rm BG}}}{dt\, dp_s}=\pi m m_D \frac{p_s}{E_s} |\psi_D(p_s)|^2
\frac{d \sigma^p}{dt}  \int^{E_{\Lambda}^{\rm max}}_{E_{\Lambda}^{\rm min}} \frac{d E_{\Lambda}}{2 |p_{\gamma}-p_{\Lambda}|} \Theta\left(-1 < \frac{m_K^2+|p_{\gamma}-p_{\Lambda}|^2+p_s^2-b^2}{2 p_s |p_{\gamma}-p_{\Lambda}|}  <1 \right) \,,
\label{eq:b10}
\end{equation}
where  $p_{\Lambda}$ is the momentum
of ${\Lambda}$ in the deuteron rest frame; $b=E_{\gamma}+m_D-E_s-E_{\Lambda}$;
$d \sigma^p /dt$
is the cross section of the $\gamma +p \to \Lambda (\Sigma)+K^+$ reaction. 
The upper and lower limits of integration over $E_{\Lambda}$ are determined by the
condition on the $n-K^+$ invariant mass (see Eq.~(\ref{eq:bg}))
\begin{equation}
E_{\Lambda}^{\rm min, max}=\left(t+m_D^2+2m_D E_{\gamma}-(M_{\Theta} \pm 0.010)^2 \right)/ (2 m_D) \,.
\label{eq:b11}
\end{equation}
This cross section is given by dashed curves in Fig.~\ref{fig:risunok5}.

If the experimental resolution in the invariant mass of the final $K^+ n$ system is 
at the level of a few MeV and statistics is high, one can also study the signal, interference
and background cross sections as functions of the invariant mass 
$M^2=(p_s+p_K)^2$. 
The resulting expressions read
\begin{eqnarray}
&&\frac{d^3 \sigma}{dt\, dp_s \,d M^2}=\frac{d \sigma}{dt\, dp_s} \frac{\Gamma^{tot} M_{\Theta}}{\pi} \frac{1}{(M^2- M_{\Theta}^2)^2+(\Gamma^{tot} M_{\Theta})^2 } \,, \nonumber\\
&&\frac{d^3 \sigma^I}{dt\, dp_s \,d M^2}=\frac{d \sigma^I}{dt\, dp_s} \frac{\Gamma^{tot} M_{\Theta}}{\pi} \frac{1}{(M^2- M_{\Theta}^2)^2+(\Gamma^{tot} M_{\Theta})^2 }+\dots \,, \nonumber\\
&&\frac{ d^3 \sigma^{{\rm BG}}}{dt\, dp_s\,d M^2}=\pi m \frac{p_s}{E_s} |\psi_D(p_s)|^2
\frac{d \sigma^p}{dt}   \frac{1}{4 |p_{\gamma}-p_{\Lambda}|} \Theta\left(-1 < \frac{m_K^2+|p_{\gamma}-p_{\Lambda}|^2+p_s^2-b^2}{2 p_s |p_{\gamma}-p_{\Lambda}|}  <1 \right) 
\label{eq:b12}
\end{eqnarray}
In the second line, the dots denote a correction which introduces a small
deviation of the interference cross section from the pure Breit-Wigner form.
The cross sections of Eq.~(\ref{eq:b12}) are presented in Fig.~\ref{fig:risunok7}.

\end{document}